\documentclass[article]{IEEEtran}
\usepackage{cite}
\usepackage[cmex10]{amsmath}
\usepackage{amssymb}
\usepackage{amsthm}
\usepackage{mathrsfs}
\usepackage{bm}
\usepackage[mathscr]{eucal}
\usepackage{amssymb,amsmath,amsthm,amsfonts,latexsym}
\usepackage{amsmath,graphicx,bm,xcolor,url}
\usepackage{graphicx}
\usepackage{latexsym}
\usepackage{CJK}
\usepackage{indentfirst}
\usepackage{geometry}
\usepackage{psfrag}
\usepackage{setspace}
\usepackage{algorithmic}
\usepackage{algorithmic, cite}
\usepackage{algorithm}
\usepackage{array}
\usepackage{mdwmath}
\usepackage{mdwtab}
\usepackage{eqparbox}
\usepackage{url}
\usepackage{epstopdf}
\usepackage{epsfig,epsf,psfrag}
\usepackage{fixltx2e}
\usepackage{verbatim}
\usepackage{textcomp}
\usepackage{psfrag} 

\usepackage{caption}
\usepackage{subcaption}

\theoremstyle{plain}

\theoremstyle{remark}

\theoremstyle{plain}

\theoremstyle{remark}

\theoremstyle{plain}

\theoremstyle{remark}

\theoremstyle{remark}

\theoremstyle{remark}

\theoremstyle{remark}

\theoremstyle{remark}

\theoremstyle{remark}

\catcode`~=11 \def\UrlSpecials{\do\~{\kern -.15em\lower .7ex\hbox{~}\kern .04em}} \catcode`~=13

\allowdisplaybreaks[4]

\newcommand{\calA}{\mathcal{A}}

\newcommand{\bg}{\mathbf{g}}

\newcommand{\bh}{\mathbf{h}}
\newcommand{\bH}{\mathbf{H}}

\newcommand{\bI}{\mathbf{I}}

\newcommand{\bq}{\mathbf{q}}

\newcommand{\bT}{\mathbf{T}}
\newcommand{\bu}{\mathbf{u}}

\newcommand{\bv}{\mathbf{v}}

\newcommand{\bx}{\mathbf{x}}

\newcommand{\by}{\mathbf{y}}

\newcommand{\bz}{\mathbf{z}}

\newcommand{\bbC}{\mathbb{C}}

\newcommand{\bbE}{\mathbb{E}}

\DeclareMathAlphabet{\mathbsf}{OT1}{cmss}{bx}{n}
\DeclareMathAlphabet{\mathssf}{OT1}{cmss}{m}{sl}

\DeclareSymbolFont{bsfletters}{OT1}{cmss}{bx}{n}
\DeclareSymbolFont{ssfletters}{OT1}{cmss}{m}{n}
\DeclareMathSymbol{\bsfGamma}{0}{bsfletters}{'000}
\DeclareMathSymbol{\ssfGamma}{0}{ssfletters}{'000}
\DeclareMathSymbol{\bsfDelta}{0}{bsfletters}{'001}
\DeclareMathSymbol{\ssfDelta}{0}{ssfletters}{'001}
\DeclareMathSymbol{\bsfTheta}{0}{bsfletters}{'002}
\DeclareMathSymbol{\ssfTheta}{0}{ssfletters}{'002}
\DeclareMathSymbol{\bsfLambda}{0}{bsfletters}{'003}
\DeclareMathSymbol{\ssfLambda}{0}{ssfletters}{'003}
\DeclareMathSymbol{\bsfXi}{0}{bsfletters}{'004}
\DeclareMathSymbol{\ssfXi}{0}{ssfletters}{'004}
\DeclareMathSymbol{\bsfPi}{0}{bsfletters}{'005}
\DeclareMathSymbol{\ssfPi}{0}{ssfletters}{'005}
\DeclareMathSymbol{\bsfSigma}{0}{bsfletters}{'006}
\DeclareMathSymbol{\ssfSigma}{0}{ssfletters}{'006}
\DeclareMathSymbol{\bsfUpsilon}{0}{bsfletters}{'007}
\DeclareMathSymbol{\ssfUpsilon}{0}{ssfletters}{'007}
\DeclareMathSymbol{\bsfPhi}{0}{bsfletters}{'010}
\DeclareMathSymbol{\ssfPhi}{0}{ssfletters}{'010}
\DeclareMathSymbol{\bsfPsi}{0}{bsfletters}{'011}
\DeclareMathSymbol{\ssfPsi}{0}{ssfletters}{'011}
\DeclareMathSymbol{\bsfOmega}{0}{bsfletters}{'012}
\DeclareMathSymbol{\ssfOmega}{0}{ssfletters}{'012}

\newcommand{\hatc}{\widehat{c}}

\newcommand{\tilbH}{\widetilde{\bH}}

\newcommand{\hats}{\widehat{s}}

\newcommand{\tilbu}{\widetilde{\bu}}

\newcommand{\tilbx}{\widetilde{\bx}}

\newcommand{\tilby}{\widetilde{\by}}

\newcommand{\barx}{\bar{x}}

\def\norm#1{\left\| #1 \right\|}
\def\norm2#1{\left\| #1 \right\|_2}
\def\norm22#1{\left\| #1 \right\|_2^2}

\DeclareMathOperator{\diag}{diag}

\newcommand{\qednew}{\nobreak \ifvmode \relax \else
      \ifdim\lastskip<1.5em \hskip-\lastskip
      \hskip1.5em plus0em minus0.5em \fi \nobreak
      \vrule height0.75em width0.5em depth0.25em\fi}

\usepackage{tabularx}
\usepackage{graphics}
\usepackage{bbm}

\usepackage{stfloats}

\title{Symbiotic Radio: Cognitive Backscattering Communications for Future Wireless Networks}

\author{Ying-Chang Liang, \emph{Fellow, IEEE}, Qianqian Zhang, Erik G. Larsson, \emph{Fellow, IEEE}, \\
and Geoffrey Ye Li, \emph{Fellow, IEEE} \\

\thanks{
Y.-C. Liang is with the Center for Intelligent Networking and Communications (CINC), University of Electronic Science and Technology of China (UESTC), Chengdu 611731, China (e-mail: liangyc@ieee.org).

Q.~Zhang is with the National Key Laboratory of Science and Technology on Communications, and the Center for Intelligent Networking and Communications (CINC), University of Electronic Science and Technology of China (UESTC), Chengdu 611731, China (e-mails: qqzhang\_kite@163.com).

E. G. Larsson is with the Department of Electrical Engineering (ISY), Linköping University, SE-581 83 Linköping, Sweden (email: erik.g.larsson@liu.se).

G. Y. Li is with the School of Electrical and Computer Engineering, Georgia Institute of Technology, Atlanta, GA 30332-0250 USA (e-mail:
liye@ece.gatech.edu).
}}
\begin{document}
\maketitle
\begin{abstract}

The heterogenous wireless services and exponentially growing traffic call for novel spectrum- and energy-efficient wireless communication technologies.
In this paper, a new technique, called symbiotic radio (SR), is proposed to exploit the benefits and address the drawbacks of \emph{cognitive radio} (CR) and \emph{ambient backscattering communications} (AmBC), leading to mutualism spectrum sharing and highly reliable backscattering communications.
In particular, the \emph{secondary transmitter} (STx) in SR transmits messages to the \emph{secondary receiver} (SRx) over the RF signals originating from the \emph{primary transmitter} (PTx) based on cognitive backscattering communications, thus the secondary system shares not only the radio spectrum, but also the power, and infrastructure with the primary system. In return, the secondary transmission provides beneficial multipath diversity to the primary system, therefore the two systems form mutualism spectrum sharing. More importantly, joint decoding is exploited at SRx to achieve highly reliable backscattering communications.
To exploit the full potential of SR, in this paper, we address three fundamental tasks in SR: (1) enhancing the backscattering link via active load; (2) achieving highly reliable communications through joint decoding; and (3) capturing PTx's RF signals using reconfigurable intelligent surfaces. Emerging applications, design challenges and open research problems will also be discussed.

\end{abstract}


\section{Introduction}



The development of digitalized society depends significantly on the advancements of wireless communication technologies. In order to further realize smart society, a variety of emerging wireless services need to be developed, such as holographic telepresence, \emph{Internet-of-Everything} (IoE), and collaborative intelligence, which will drive the dramatic growth in data traffic, massive access, and versatile applications.
According to the report from \emph{International Telecommunication Union} (ITU) \cite{imt2015}, the global mobile traffic will continue to grow exponentially and the overall mobile data traffic is estimated to reach over $5000$ EB per month in 2030.
The device density will increase substantially to hundreds of devices per cubic meter and more than 125 billion mobile devices worldwide will be attached to the wireless networks by 2030 \cite{matti2019key}. According to a study in \cite{e2016Identification}, even just to support \emph{Internet-of-Things} (IoT) for healthcare, utility, and motorway use, 76 GHz spectrum is required if dedicated spectrum is allocated to each service. However, most of the radio spectrum has been allocated to the existing services and applications, resulting in severe spectrum scarcity problem.
The explosive growth in data traffic and device density calls for novel technologies to enhance the spectrum efficiency as well as energy-efficiency of wireless communications.

In the past two decades, \emph{cognitive radio} (CR) technology has been extensively studied with the aim to enhance the utilization efficiency of the radio spectrum \cite{mitola1999cognitive, haykin2005cognitive, liang2011cognitive, qin201920}. This is achieved through introducing secondary spectrum access to the assigned spectrum. In particular, the secondary user in a CR system is allowed to access the radio spectrum assigned to the primary user in an opportunistic or spectrum sharing manner. The opportunistic access is assisted by spectrum sensing technologies, while in spectrum sharing model, the secondary transmission from the \emph{second transmitter} (STx) to the \emph{second receiver} (SRx) shares the same spectrum with the primary transmission from the \emph{primary transmitter} (PTx) to the \emph{primary receiver} (PRx). This is achieved by predicting the interference level from STx to PRx to ensure that the caused interference to the primary transmission is below a tolerable threshold. Despite the notable advancements, the primary and secondary transmissions in spectrum sharing model always interfere with each other, which limits the enhancement of the spectrum efficiency.

On the other hand, conventional transmitter design uses power-consuming active components, including oscillators, up-converters, and power amplifiers. Such active radio technology involves high power consumption, shortens battery life, and limits the development of the emerging services in future wireless networks \cite{report2020from}. Recently, backscattering radio technology has been exploited to design wireless transmitters without requiring active components, thus the power consumption of the transmitter can be greatly reduced. One example of passive radio technology is \emph{ambient backscatter communications} (AmBC) \cite{liu2013ambient}, in which the STx embeds its message over the ambient \emph{radio frequency} (RF) signals using backscattering modulation \cite{Boyer2014Backscatter}. The ambient RF signals can be TV signal, WiFi signal, cellular signal, etc. In particular, by periodically varying the load impedance at the STx, the reflected signal from the STx contains different reflection coefficients representing the transmitted symbols of the STx. Thus through detecting the reflection coefficients, the SRx can decode the messages transmitted from STx. For recent development on AmBC, please refer to, e.g., \cite{van2018ambient}.

In AmBC, the secondary and ambient communications form a special spectrum sharing relation, and the design objective at SRx is to recover the messages from STx only. Due to the ambient nature of the RF sources, non-coherent detectors have been extensively studied. For example, differential demodulation is proposed in \cite{liu2013ambient}, and energy detection has been considered in, e.g., \cite{liu2013ambient,wang2016ambient,qian2017noncoherent,qian2017semi,qian2018iot,liu2017coding}. These non-coherent receivers suffer from strong direct link interference from the RF source, which degrades the detection performance significantly. While there are proposals to avoid the direct link interference using frequency shifting \cite{Iyery_AmBC-frquency-fisf2016,zhang_AmBC-frquency-fisf2016,elmossallamy2019noncoherent} or to cancel out such interference using specific waveform feature of the RF source \cite{yang2016backscatter,yang2018modulation,nguyen2019signal,zhang2018interference}, the non-coherent nature of the receivers makes it difficult for AmBC to achieve highly reliable backscattering communications.

In this paper, a new technique, called \emph{symbiotic radio} (SR), is proposed to exploit the benefits of CR and AmBC, and address the drawbacks of these two techniques effectively. Similar to CR, SR consists of two spectrum sharing systems, the primary system and secondary system. However, SR achieves mutually beneficial spectrum sharing rather than interfering spectrum sharing in CR. As compared to AmBC, SR achieves highly reliable backscattering communications through joint decoding. Thus, SR is also termed cognitive backscattering communications, which achieves enhanced spectrum- and energy-efficiency for wireless networks.

This paper is organized as follows. The SR basics and backscattering communications for SR are presented in Sections II and III, respectively. Then, in Sections IV and V, the transceiver design, and resource allocation schemes for SR are presented, respectively. \emph{Reconfigurable intelligent surface} (RIS)-assisted SR and full-duplex SR are addressed in Section VI and VII, respectively. Emerging applications, design challenges, and open research problems will be discussed in the rest parts of the paper.

The list of abbreviations commonly appeared in this paper is given in Table \ref{tab2}.

\begin{table}[htbp]
\caption{List of Abbreviations}
\begin{center}
\begin{tabular}{|l|p{5.2cm}|}
\hline
{\textbf{Abbreviation}}& \textbf{Description}\\
\hline
{AmBC}&{Ambient Backscatter Communications}\\
\hline
{AWGN}&{Additive White Gaussian Noise}\\
\hline
{BER}&{Bit-Error-Rate}\\
\hline
{BPSK}&{Binary Phase Shift Keying}\\
\hline
{CR}&{Cognitive Radio}\\
\hline
{DoA}&{Direction-of-Arrival}\\
\hline
{FDSR}&{Full-Duplex SR}\\
\hline
{IoE}&{Internet-of-Everything}\\
\hline
{IoT}&{Internet-of-Things}\\
\hline
{ITU}&{International Telecommunication Union}\\
\hline
{MISO}&{Multiple-Input-Single-Output }\\
\hline
{MMSE}&{Minimum Mean-Squared-Error}\\
\hline
{MRC}&{Maximum-Ratio-Combining}\\
\hline
{OFDM}&{Orthogonal Frequency Division Multiplexing}\\
\hline
{PRx}&{Primary Receiver}\\
\hline
{PTx}&{Primary Transmitter}\\
\hline
{QAM}&{Quadrature Amplitude Modulation}\\
\hline
{RF}&{Radio Frequency}\\
\hline
{RIS}&{Reconfigurable Intelligent Surface}\\
\hline
{RSR}&{RIS-assisted SR}\\
\hline
{SISO}&{Single-Input-Single-Output}\\
\hline
{SNR}&{Signal-to-Noise-Ratio}\\
\hline
{SR}&{Symbiotic Radio}\\
\hline
{SRx}&{Second Receiver}\\
\hline
{STx}&{Second Transmitter}\\
\hline
{ZF}&{Zero-Forcing}\\
\hline
\end{tabular}
\label{tab2}
\end{center}
\end{table}

\section{Symbiotic Radio Basics}

\begin{figure}[t]
    \centering
    \begin{subfigure}{0.8\linewidth}
     \centering
    \includegraphics[width=0.8\columnwidth]{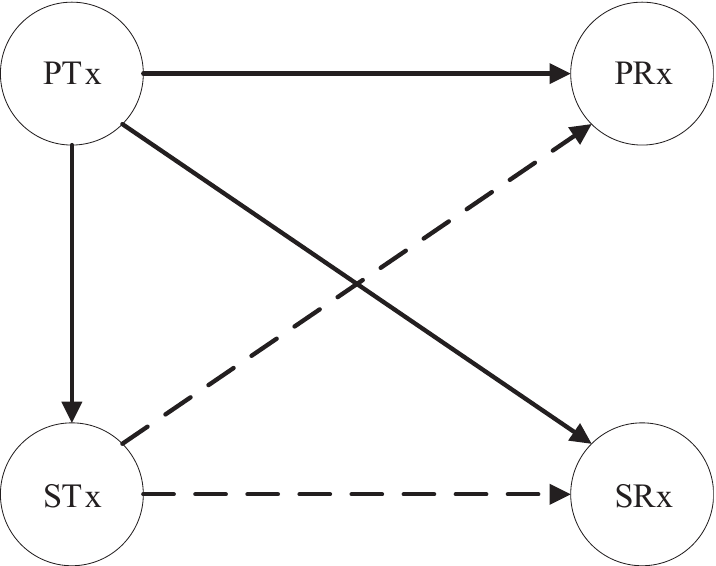}
    \caption{System model for SR.}
    \end{subfigure}
    \begin{subfigure}{0.8\linewidth}
    \centering
    \includegraphics[width=0.8\columnwidth]{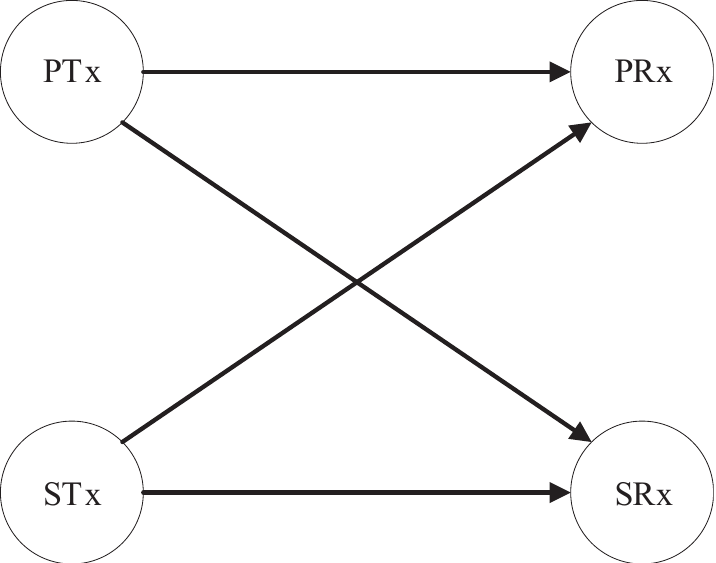}
   \caption{System model for CR.}
    \end{subfigure}
    \caption{(a) System model for SR: PTx uses active radio to transmit messages to PRx, and STx exploits backscattering radio to transmit messages to SRx riding over the RF signals from PTx; and (b) System model for CR: PTx transmits messages to PRx, and STx transmits messages to SRx by sharing the same radio spectrum with the primary system. STx refrains its transmit power to ensure that the caused interference level to PRx is below a tolerable threshold. Solid lines denote transmissions from active radios, while dash lines denote reflections from backscattering radios.}

    \label{fig:system}
\end{figure}

SR is a cognitive backscattering communication system that exploits the benefits and addresses the drawbacks of CR and AmBC. The system model for a basic SR is shown in Fig.~\ref{fig:system}(a), which consists of two systems, the primary system and secondary system. Different from CR that uses power-consuming active RF chains at both PTx and STx shown in Fig.~\ref{fig:system}(b), the SR uses backscattering radio technology to support the secondary transmission from STx to SRx, which greatly reduces the power consumption. In particular, STx transmits its messages to SRx over the RF signals received from PTx by varying the reflection coefficients, thus the secondary system shares the spectrum, energy, and infrastructure of the primary system.

More importantly, the two systems in SR operate in a collaborative manner in both the transmitter and receiver sides. By doing so, joint decoding is exploited at the SRx to decode both primary and secondary messages, thus avoiding the interference issue, and achieving highly reliable backscattering communications. In return, the secondary system provides multipath diversity to the primary system, yielding mutual benefits between the two spectrum sharing systems. Therefore, SR is also called cognitive backscattering communications which achieves two challenging goals for wireless communications:

\begin{itemize}

\item[(1)] enhanced spectrum efficiency using mutualism spectrum sharing;
\item[(2)] enhanced energy efficiency through highly reliable backscattering communications.

\end{itemize}

The beneficial effect of the backscattering radio to primary system was first observed in \cite{YangLiangZhangICC17,yang2018cooperative} and \cite{liu2018backscatter}, and the symbiotic relation between the two systems was revealed in \cite{guo2019cooperative}. Later on, several studies on SR were published in, e.g.,  \cite{zhang2019backscatter, long2019full, guo2019resource, long2019symbiotic}. SR has also been recently considered as a potential solution for the \emph{six generation} (6G) wireless networks, especially as an energy- and spectrum-efficient massive access technology \cite{nawaz2020non,bariah2020prospective,chen2020vision,8820755}.

Despite the advantages of SR, several fundamental challenges need to be tackled in order to fully exploit the potentials of SR.
From Fig.~\ref{fig:system}(a), the backscattering link from PTx-STx-SRx suffers from double fading, thus its strength is much weaker than that of the direct link from PTx to SRx. It is thus important to enhance the backscattering link such that the mutual benefits to the primary and secondary systems can be improved. On the other hand, the backscattering radios can backscatter all ambient signals within a specific frequency band, thus it is essential to capture which RF signals the secondary system wants to ride on. Finally, effective transceiver designs are required in order to achieve highly reliable backscattering communications.

In this paper, we will address the following issues related to SR:

\begin{itemize}

\item[(1)] enhancing the backscattering link via active load and/or multiple antenna elements;
\item[(2)] achieving highly reliable communications through joint decoding;
\item[(3)] capturing PTx's RF signals using RIS.

\end{itemize}

When the STx is equipped with RIS, this system is called \emph{RIS-assisted SR} (RSR). Note in some cases, the PRx and SRx in Fig.~\ref{fig:system}(a) can be the same node, which is to decode the primary and secondary messages simultaneously. In addition, when the PTx has full duplex functionality, it can be used to decode the messages from STx, and this model is called \emph{full-duplex SR} (FDSR).

\section{Backscattering Communication for SR}\label{sec:backscattering}

 In this section, we will introduce the backscattering principles, backscattering modulation, and schemes to enhance the backscattering link via active load, which will serve as the basis for SR.

\subsection{Backscattering Principles}\label{sec:backscatter}
\begin{figure}
  \centering
  \includegraphics[width=.5\columnwidth] {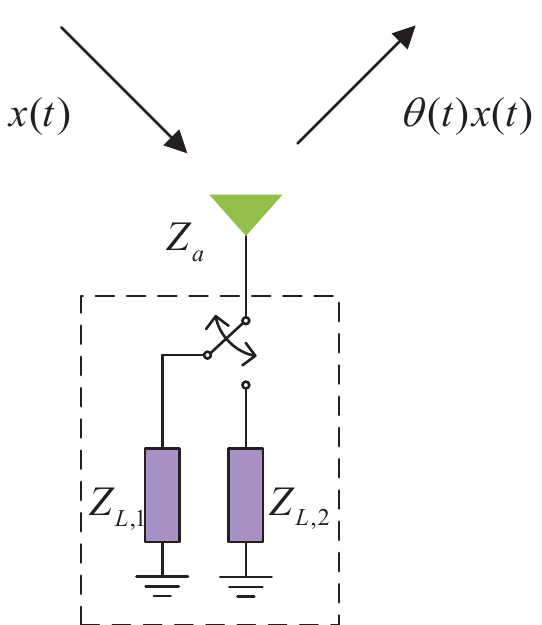}
  \caption{Basic backscattering circuits.}
  \label{fig:backscattering}
  \end{figure}

The basic circuit for backscattering radios is shown in Fig.~\ref{fig:backscattering}. Suppose that $s(t)$ is baseband signal transmitted from PTx, the RF output from the PTx antenna is $x_{0}(t) = {\rm{Re}} \{ \sqrt{p} s(t)e^{j 2\pi f_{c}t}\}$, where $f_{c}$ is the carrier frequency.  Let $x(t)$ be the input to the backscattering circuit at STx, and $\theta(t)$ be the reflection coefficient, then signal reflected from the backscattering circuit is $\theta(t) x(t)$.

Let $Z_a$ and $Z_{L,i}$ be the antenna and load impedances of the backscattering circuits at STx, respectively. From the antenna scattering theory \cite{Hansen1989}, the electric field of the reflected signal can be decomposed into two components, the structural mode backscattering component, and the antenna mode backscattering component.
The structural mode backscattering component can be accounted into part of the environmental multipath \cite{Thomas2012}, while the antenna mode backscattering component is determined by the mismatch between the antenna and load impedances, yielding the reflection coefficient given by
\begin{align}
\Gamma_i =\frac{Z_{L,i}-Z_a^{*}}{Z_{L,i}+Z_a}\label{eq:scatter1}.
\end{align}
Note when $Z_{L,i} = Z_a^{*}$, $\Gamma_{i} = 0$. Thus, by changing the value of the load impedance with a switcher, the STx can generate the following time-varying reflection coefficient
\begin{align}
  \theta (t) =\Gamma_{i}, \label{eq:scatter1}
\end{align}
if load $Z_{L,i}$ is switched on at time instant $t$.



\subsection{Backscattering Modulation}
By changing the load impedance periodically, the STx can generate different reflection coefficients, which represent the information to be sent to the SRx. The key issue for backscattering modulation is to design an appropriate reflection coefficient set corresponding to the signal constellation set $\calA_{\sf c}$.
Basically, the number of the load impedance states attributes to the modulation order. Taking \emph{binary phase shift keying} (BPSK) as an example, the STx only needs two load impedance states to represent symbols '-1' and '1'. In order to improve the transmission efficiency, high order modulation schemes, like \emph{quadrature amplitude modulation} (QAM), can be developed. The relation between the complex constellation point $c_i \in\calA_{\sf c}$ and the reflection coefficient $\Gamma_i$ is given by \cite{Thomas2012Quadrature}
\begin{equation}\label{eq:backCon}
\Gamma_i = \alpha \cdot \frac{c_i}{|c|_{max}},
\end{equation}
where $\alpha$ accounts for the reflection efficiency of the circuits \cite{Boyer2014Backscatter} and its value is generally $0\leq\alpha\leq1$ with the passive load, and $|c|_{max} = \max_{c\in\calA_{\sf c}}|c|$ is the largest amplitude of the constellation points, which is presented in the denominator due to the limitation that the passive reflection coefficient cannot be greater than unity. Without loss of generality, we normalize the constellation points such that $|c|_{max}=1$. Once the reflection coefficient $\Gamma_i$ is given, the corresponding load impedance $Z_{L,i}$ is chosen as
\begin{equation}\label{eq:loadIm}
Z_{L,i}=\frac{Z_\mathrm{a}^{*}+\Gamma_i Z_\mathrm{a}}{1-\Gamma_i}.
\end{equation}

\subsection{Enhancing Backscattering Link}

From Fig.~\label{fig:trans}(a), the backscattering link suffers from double fading, and thus the strength of the backscattering link signal in SR is much
weaker than that of the direct link signal, which will limit the performance of the secondary system and the improvement to the primary system.
Typically, there are two solutions to enhance the strength of the backscattering link. One is to amplify the backscattering link by using an active load at the STx. The other is to introduce multiple antenna elements or called RIS at the STx.
The use of RIS in STx can not only enhance the backscattering link, but also capture the RF signals from desired sources using passive beamforming, and its details will be discussed in Section \ref{sec:RIS}. Here, we mainly describe the amplification principle based on the backscattering technology.


Recent advances in backscattering communications show that the STx can take advantages of the active load, whose resistance is negative, to amplify the incident signal \cite{Amato2018}. Specifically, provided that the impedance of the active load is $Z_{L,i}=-R_{L,i}+jX_{L,i},\quad R_{L,i}>0$, and the antenna impedance is $Z_{A}=R_{A}+jX_A,~R_{A}>0$, the reflection coefficient is then characterized with
\begin{equation}\label{eq:AL_RE}
  |\Gamma_{i}|^2 =\frac{\left(R_{L,i}+R_A\right)^2+\left(X_{L,i}+X_A\right)^2}{\left(R_{L,i}-R_A\right)^2+\left(X_{L,i}+X_A\right)^2}>1,
\end{equation}
which implies that the STx can amplify the incident signal. Such signal amplification however needs additional biasing source to support the active load, which can be realized by Tunnel diodes \cite{1965APL,Amato2018,Amato2018a,Khaledian2019} or CMOS technology \cite{Bousquet20124}.

\section{Transceiver Design for SR} \label{sec:transceiver}

\begin{figure}[t]
    \centering
    \begin{subfigure}{1.0\linewidth}
    \centering
    \includegraphics[width=0.9\columnwidth]{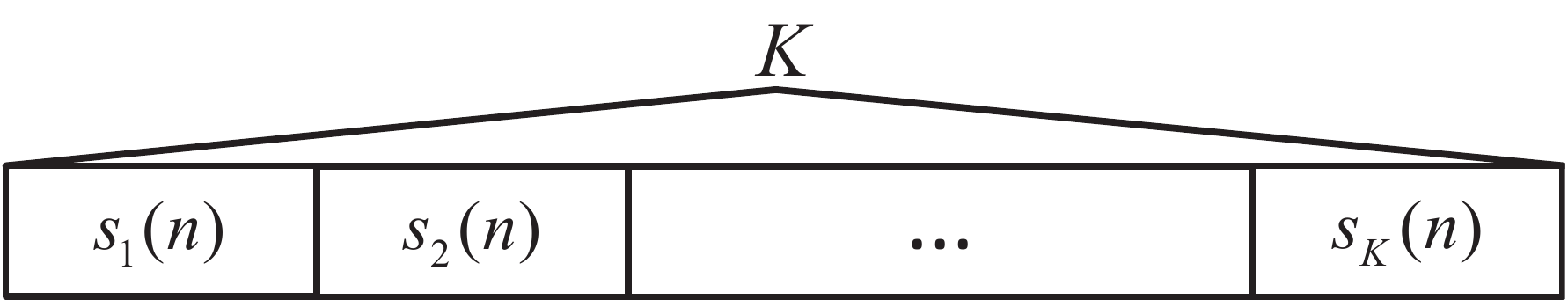}
    \caption{Primary transmission.}
    \end{subfigure}
    \begin{subfigure}{1.0\linewidth}
    \centering
    \includegraphics[width=0.9\columnwidth]{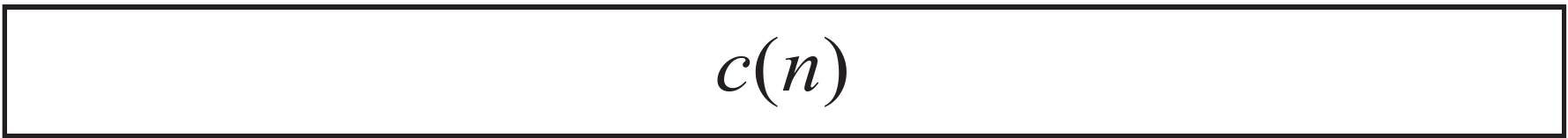}
    \caption{Secondary transmission.}
    \end{subfigure}
    \caption{Illustration for SR transmissions: (a) Primary signal $s_k(n)$; (b) Secondary signal $c(n)$. One secondary symbol period covers $K$ primary symbol periods.}

    \label{fig:trans}
    \vspace{-0.5em}
\end{figure}

As with every communication system, the receivers are required to recover the messages transmitted from the transmitters. In the proposed SR system, the PRx and SRx are supposed to recover the primary and secondary information messages, respectively. However, in order to fully exploit the potentials of SR, we use joint decoding to recover both primary and secondary messages in each receiver. To do so, we will first provide detailed description on the signal models and transceiver design schemes. We will also compare SR with CR, and SR with AmBC in details.


\subsection{Signal Models}

Denote $T_s$ and $T_c$ as the symbol periods for the primary transmission and the secondary transmission, respectively. In SR, we assume that each secondary symbol period covers $K$ ($K$ is an integer) primary symbol periods, i.e., $T_{c} = K T_{s}$, as shown in Fig.~\ref{fig:trans}. Denote $c(n)\in \calA_{\sf c}$ as the $n$th symbol of the secondary transmission and $s_{k}(n)\in\calA_{\sf s}$ the $k$th primary symbol within the $n$th secondary symbol period, where $k = 1,\cdots,K$.
Note that synchronization between $s_{k}(n)$ and $c(n)$ is required for $K=1$ to avoid spectrum growth, but for large $K$, the spectrum growth due to asynchronous transmission of $s_{k}(n)$ and $c(n)$ becomes negligible \cite{guo2019cognitive,ruttik2018does}.

We assume that the SRx has $M_r$ antennas and the channel remains unchanged during a secondary data frame but may vary from one frame to another.
Denote by $h_{m,1}$ the channel coefficient from the PTx to the $m$-th receive antenna at the SRx, by $l$ the channel coefficient from the PTx to the STx, and by $g_m$ the channel coefficient from the STx to the $m$-th receive antenna at the SRx.
Let $p$ be the average transmit power at the PTx. The received signal at the SRx can be written as
\begin{align}
\by_k(n) = {\sqrt{p} \bh_1 s_k(n)}+ {\sqrt{p} \bh_2 s_k(n) c(n)} + \bu_k(n), \label{eq:received}
\end{align}
for $k=1,\cdots,K$, where $\bh_1 = [h_{1,1}, \cdots, h_{M_r,1}]^T$, $\bg = [g_1,\cdots, g_{M_r}]^T$, $\bh_2 = \alpha l\bg$, and $\bu_{k}(n) \sim \mathcal{CN}(0,\sigma{^2} \bI)$. In \eqref{eq:received}, the first and second terms of the right hand side represent the direct link and backscattering link components, respectively.

 Similarly, the received signal at the PRx can be written as
\begin{align}
\bz_k(n) = {\sqrt{p} \mathbf f_1 s_k(n)}+ {\sqrt{p} \mathbf f_2 s_k(n) c(n)} + \bv_k(n), \label{eq:receivedPR}
\end{align}
for $k=1,\cdots,K$, where $\mathbf f_1$ is the channel information from the PTx to the PRx, $\bq$ is the channel information from the STx to the PRx, $\mathbf f_2 = \alpha l\bq$, and $\bv_{k}(n) \sim \mathcal{CN}(0,\sigma{^2} \bI)$.
Due to the symmetry of the transmission structure in \eqref{eq:received} and \eqref{eq:receivedPR}, we consider the signal detection at SRx only. The extension to the signal detection at PRx is straightforward.

\subsection{Coherent Receiver Design}\label{sec:coherent}

When pilot symbols for both primary and secondary transmissions are jointly designed and transmitted, the channel responses $\bh_1$ and $\bh_2$ can be estimated at the SRx. Coherent receivers \cite{yang2018cooperative}, such as optimal \emph{maximum-likelihood} (ML) detector, linear detector, and \emph{successive interference cancellation} (SIC)-based detector, can be designed to jointly recover the primary and secondary signals.


\subsubsection{ML Detector}


Let $c(n) \in \calA_{\sf c}$ and $s_{k}(n) \in \calA_{\sf s}$, the ML detector recovers $\hat \bx_{ml}(n) = [\hat s_1(n),\cdots,\hat s_{K}(n),\hat c(n)]^T$ by searching the candidate with minimum squared error, i.e.,
\begin{align} \label{eq:ML_flat}
 &\hat \bx_{ml}(n)\!=\!  \nonumber\\
 &\underset{\substack{c(n) \in \calA_{\sf c}, \\ s_k(n) \in \calA_{\sf s}, \forall k}}{\arg \min} \! \sum \limits_{k=1}^{K} \! \left\| \by_k(n) \!-\!\sqrt{p}\bh_1 s_k(n)  \!-\! \sqrt{p}\bh_2 s_k(n)c(n) \right\|^2.
\end{align}

Note the ML detector only requires to search $K|\calA_{\sf c}||\calA_{\sf s}|$ possible symbol sets. This is achieved by the following steps. Firstly, we apply spatial domain \emph{maximum-ratio-combining} (MRC) to decode $s_k(n)$ conditioned on each $c(n)$ candidate in $\calA_{\sf c}$ and obtain $\hats_k(n)|_{c(n)}$.
Then, we obtain the estimated $\hatc(n)$ with the minimal squared error by searching all the conditional estimates $\hats_k(n)|_{c(n)}$ for each $c(n)$.
Finally, we obtain the estimated $\hats_k(n) = \hats_k(n) |_{\hatc(n)}$ corresponding to the estimated $\hatc(n)$.

\subsubsection{Linear Detectors}\label{sec:linear}

The received signal at PRx can be written as $ \by_k(n) = \sqrt{p}\bH \bx_k(n) + \bu_k(n)$, where $\bH = [\bh_1,\bh_2]$, $\bx_k(n) = [s_k(n), s_k(n)c(n)]^T$. Linear detectors, such as MRC, \emph{zero-forcing} (ZF), and \emph{minimum mean-squared-error} (MMSE) detectors, can be used to extract $\bx_k(n)$, and then $s_k(n)$ and $c(n)$ are recovered from $\bx_k(n)$.
Specifically, the received signal can be simplified as
 \begin{align}\label{eq:rx_sig_mtx_composite}
   \tilby(n) = \sqrt{p}\tilbH \tilbx(n) + \tilbu(n),
 \end{align}
 where $\tilby(n) =[\by_1(n),\ldots,\by_{K(n)}]^T \in \bbC^{M_rK \times 1}$, $\tilbH=\diag \{\bH, \ \ldots, \ \bH\} \in \bbC^{M_rK \times 2K}$, $\tilbx(n) =[\bx_1(n),\ldots,\bx_{K}(n)]^T \in \bbC^{2K \times 1}$, and $\tilbu(n) =[\bu_1(n),\ldots,\bu_{K}(n)]^T \in \bbC^{M_rK \times 1}$.
 Then, the linear detectors extract $\tilbx(n)$ by
 \begin{align}
   {\overline{\bx}}(n) = \bT \tilby(n), \label{eq:linear_receiver}
 \end{align}
 where $\bT = \diag \{\bT_1; \ldots, \bT_{K}\} \in \bbC^{2K \times M_rK}$, and the corresponding $\bT_k$ is given as follows:
 \begin{align}
 \bT_k  \!=\! \left\{ \!
 \begin{array}{cl}
 \left[ \frac{\bh_1^H }{\left\| \bh_1 \right\|^2}; \frac{\bh_2^H }{\left\| \bh_2 \right\|^2} \right], &\text{for \ \ MRC}  \\
     \left( \bH^H \bH  \right)^{-1} \bH^H , &\text{for \ \ ZF, }  \\
     \left( \bH^H  \bH  + \frac{\sigma^2}{p} \bI_2 \right)^{-1} \bH^H , &\text{for \ \ MMSE},
   \end{array}
   \right.
 \end{align}
 where $\bI_2$ is a two-dimensional identity matrix.
 After the linear detection, the primary and secondary symbols can be recovered by
 \begin{align}
   \hats_k(n)& \!=\! \underset{s_k(n) \in \calA_{\sf s}}{\arg \min} \left | s_k(n)  \!-\! \barx_{2k-1}(n)\right |, \forall k \label{eq:sn_est}\\
   \hatc(n)&=\underset{c(n) \in \calA_{\sf c}}{\arg \min} \sum \limits_{k=1}^{K} \left | c(n) - \frac{\barx_{2k}(n)}{\hats_k(n)}\right |,
 \end{align}
 where $\barx_{k}(n)$ is the $k$-th element of ${\overline{\bx}}(n)$.
Compared with the ML detector, the linear detector has lower complexity at the cost of performance degradation.

\subsubsection{SIC-Based Detectors}

SIC technique can be used to decode $s_k(n)$ and $c(n)$. Firstly, the primary symbol $s_k(n)$ is estimated by using linear detector. Then the receiver subtracts the direct link signal with the estimated $\hats_k(n)$ and detects the secondary symbol $c(n)$ by the MRC estimator. Finally, the receiver re-estimates primary symbol $s_k(n)$ using MRC by treating estimated $\hatc(n)$ as a part of channel.
Compared with the linear detectors, the SIC-based detectors have better \emph{bit-error-rate} (BER) performance at the cost of a slightly higher computational complexity.

In \cite{yang2018cooperative}, various types of detectors mentioned above are studied for flat fading channels and frequency-selective fading channels. For both kinds of channels, the BER expressions for all detectors are derived in closed forms.
Extensive numerical results have shown that the existence of the secondary transmission can enhance the BER performance of the primary transmission due to the additional multipath provided by the backscattering link. Meanwhile, the BER performance of both $s_k(n)$ and $c(n)$ improves as $K$ increases. Specifically, the BER performance of $c(n)$ achieves around $3$ dB \emph{signal-to-noise-ratio} (SNR) gain when $K$ increases by two times due to the spreading gain for the secondary transmission.

\subsection{Semi-blind Receiver Design}

When the SRx knows the STx pilots only and partial knowledge about the primary transmission is available, semi-blind receiver can be designed to recover the secondary message.
Particularly, machine-learning detectors have been designed to recover the secondary message by using clustering frameworks \cite{zhang2019constellation}.
The constellation learning-based signal detection scheme uses the constellation characteristic of the primary symbol to decode $c(n)$. The received signals, $\by_k(n)$, naturally fall into clusters, which can be divided into two groups, corresponding to the secondary symbols ``-1'' or ``1''.  Thus, only two secondary labels are sent before secondary data transmission to map the two groups into the secondary symbols. It is worth noting that the cluster centroids can be represented by a small number of parameters, which can be learned using small sets of data.
In addition, blind classification methods can be used to by SRx to recognize the modulation scheme of the primary system \cite{tian2018modulation}.


In \cite{wang2019machine}, classification via $k$-nearest neighbors is used to recover the STx symbols. The strong direct link signal is firstly eliminated through projecting the received signal at the SRx into its orthogonal space using the estimated \emph{direction-of-arrival} (DoA) information of the direct link signal. Then proper beamformer at the SRx is designed to construct a test statistic, which is classified to recover the transmitted secondary symbol by using the $k$-nearest neighbors classification algorithm.

\subsection{Comparison of SR with CR}
\label{sec:CR}

CR is an enabling technology to support efficient utilization of the radio spectrum, in which the secondary user is allowed to reuse the frequency bands assigned to the licensed primary user in an opportunistic or spectrum sharing manner \cite{mitola1999cognitive,haykin2005cognitive, liang2011cognitive, qin201920}.

For the opportunistic manner of CR, the secondary user carries out spectrum sensing to detect the spectrum holes where the assigned frequency bands are not being utilized at a particular time by the primary user. After sensing the spectrum hole, the STx transmits messages to the SRx on the identified spectrum hole and simultaneously frequently monitors the operating spectrum to ensure that the PTx is not active. Once the PTx is active, the secondary user needs to vacate the operating spectrum for the primary transmission. The the activity model of the primary users can be used to assist for such operation \cite{aborahama2019stochastic}. Such model can be used to describe the coexistence between LTE and WiFi spectrum sharing system \cite{huang2019achieving}.


For the spectrum sharing manner of CR, as shown in Fig.~\ref{fig:system}(b), the secondary transmission coexists with the primary transmission at the same time, as long as the caused interference to the primary transmission is below a tolerable threshold.
To calculate the caused interference at the the PRx, the STx is required to predict the channel information from the STx to the PRx.
Moreover, the STx needs to know the tolerable threshold at the PRx to protect the primary transmission.
With the above knowledge, power control is conducted at the STx to guarantee that there is no excessive interference to the PRx.
The power control problem under either \emph{additive white Gaussian noise} (AWGN) or fading channels with either the peak or the average interference power constraints are studied in \cite{ghasemi2007fundamental,musavian2009capacity,Kang2009Optimal}. 
The iterative water filling algorithm is commonly used to solve the optimal power allocation problem, in which the large gain channels are allocated with high power while the small gain channels are allocated with low power or no power.

Nevertheless, the primary and secondary transmissions in spectrum sharing model always interfere with each other, which limits the enhancement of the spectrum efficiency for CR. The secondary transmission in SR however can benefit the primary system due to the introduction of the multipath diversity.
On the other hand, both PTx and STx in CR are with power-consuming active RF chains while SR provides power-dimensional sharing and infrastructure-dimensional sharing through backscattering radios in addition to the spectrum sharing.
Therefore, compared with the CR technology, the SR technology can achieve mutual benefits and multiple dimensional resource sharing with high spectrum-, energy- and cost-efficiency.

\subsection{Comparison of SR with AmBC}
\label{sec:AmBC}

In AmBC, the primary and secondary systems are independently designed, thus the secondary system has no information about the primary system, such as modulation scheme, pilots, frame structure, etc. As such, the channel state information is difficult to acquire and noncoherent detection, such as energy detection that just classifies the average received energy into several categories, is commonly used to recover the secondary symbols \cite{liu2013ambient,wang2016ambient,qian2017noncoherent,qian2017semi,qian2018iot,liu2017coding,tao2018symbol}.

In \cite{liu2013ambient}, the differential coding is used at the STx to avoid the channel estimation. The SRx firstly detects the STx transmitted symbols with energy detector and then uses the differential decoding to recover the transmitted information. The BER performance of the energy detector with differential coding under the signal-antenna PRx is analyzed in \cite{wang2016ambient}.
In \cite{qian2017noncoherent}, the joint energy detection is used to recover the STx information, in which the differential coding characteristics is combined into the energy detector for better performance. In \cite{qian2017semi}, the pilots are transmitted at the STx to assist symbol recovery based on energy detection instead of using differential coding.
To achieve high data rate of the secondary transmission, the high-order modulation, $M$-PSK, is employed for backscattering at the STx in \cite{qian2018iot} and the optimal multilevel energy detector is used to recover the STx messages. Moreover, in \cite{liu2017coding}, the STx transmits information by adjusting three states: positive and negative phase backscatter, and non-backscatter. Then, the corresponding detector at the SRx is designed to extract STx information. Manchester coding and differential Manchester coding are adopted at the STx in \cite{tao2018symbol} to perform reliable detection and the corresponding detectors are proposed to recover the STx information.

Despite the above notable receiver design for AmBC, the energy detector suffers from a large performance loss since it treats the direct link signal as undesired interference. There are some proposals to avoid the direct link interference or to cancel out such interference.
Specifically, in \cite{Iyery_AmBC-frquency-fisf2016,zhang_AmBC-frquency-fisf2016,elmossallamy2019noncoherent}, the STx shifts the primary signal to an adjacent non-overlapping frequency band to avoid the direct link interference while this technique produces two sidebands, which may interfere with other communications.
Waveform design at the STx is applied in \cite{yang2018modulation,yang2016backscatter,nguyen2019signal,zhang2018interference} to cancel out the direct link interference by exploiting the cyclic prefix structure when the PTx transmits \emph{orthogonal frequency division multiplexing} (OFDM) signals.

While the proposals to avoid the direct link interference or to cancel out such interference can enhance the receiver performance, the non-coherent nature of the receivers makes it difficult for AmBC to achieve highly reliable backscattering communications.
However, in SR, the primary and secondary systems work in collaborative manner, through which not only the PTx and the STx can be jointly designed, but also each receiver can jointly decode the primary and secondary messages.
As a result, the existence of secondary transmission in SR can in return enhance the performance of the primary system.

To conclude, the detailed comparison among CR, AmBC, and SR is summarized in Table \ref{tab1}.

\begin{table}[htbp]
\caption{Comparison of SR with CR and AmBC}
\begin{center}
\begin{tabular}{|p{1.8cm}|p{1.6cm}|p{1.6cm}|p{1.6cm}|}
\hline
{Technology}& {CR}& {AmBC}& {SR}\\
\hline
{Operating mechanism}&{Active}&{Backscatter-ing}&{Backscatter-ing}\\
\hline
{Spectrum sharing}&{Yes}&{Yes}&{Yes}\\
\hline
{Power sharing}&{No}&{Yes}&{Yes}\\
\hline
{Transmit collaboration}&{No}&{No}&{Yes}\\
\hline
{Joint decoding}&{No}&{No}&{Yes}\\
\hline
{Relationship}&{Interference}&{Interference}&{Mutualism}\\
\hline
\end{tabular}
\label{tab1}
\end{center}
\end{table}

\section{Resource Allocation for SR}
\label{sec:Resource_allocation}

Resource allocation at the relevant nodes in SR plays an important role to fulfill the desired performance requirements for both primary and secondary transmissions. In this section, we first describe the achievable rates of the primary and secondary systems, based on which, we then address the resource allocation schemes for different setups, covering single-antenna PTx, multiple-antenna PTx, and full-duplex STx.

\subsection{Achievable Rates}

From \eqref{eq:receivedPR}, the primary symbol, $s_k(n)$, is in both the direct and backscattering links, thus the receiver can treat the backscattering link as an additional path when decoding the primary symbols. As such, the achievable rate $R_{s}$ of the primary system satisfies
\begin{equation}\label{equ:B_SBN_h0}
  R_{s} \leq \mathbb{E}_c\left[ {{{\log }_2} \left (1 + \frac{p \left \|\mathbf f_1+c\mathbf f_2 \right \|^2}{\sigma ^2} \right )} \right].
\end{equation}
This upper bound is achieved when $c(n)$ is perfectly decoded. For large $K$, such perfect decoding for $c(n)$ becomes possible due to the spreading gain. When $K = 1$, however, it is a challenging task to perfectly decode the secondary signal $c(n)$ at the receiver. In this case, a worst situation happens when the backscattering link signal is thoroughly treated as interference when decoding the primary symbols. Thus we have
\begin{equation}\label{equ:B_SBN_h1}
  R_{s} \geq  \log_2 \left (1 + \frac{p \left \|\mathbf f_1\right \|^2}{p\left \|\mathbf f_2\right \|^2 +\sigma ^2} \right ).
\end{equation}
For the secondary system, $s_{k}(n)$ is firstly decoded and then the SRx cancels the interference from the direct link signal. Thus when decoding $c(n)$, the primary signal $s_k(n)$ can be viewed as a spread-spectrum code with length $K$ \cite{yang2018modulation}. For $K = 1$,
the achievable rate of the secondary system is given by \cite{long2019symbiotic}
\begin{equation}\label{equ:para}
R_{c}=\mathbb E_{s_k(n)}\left[{{{\log }_2} \left (1 + \frac{ p \left \| {\bh}_2s_k(n)\right \|^2}{\sigma^2} \right )}\right].
\end{equation}
For large $K$, the achievable rate becomes \cite{long2019symbiotic}
\begin{equation}\label{equ:para}
R_{c}=\frac{1}{K}{{{\log }_2} \left (1 + \frac{K  p \left \| {\bh}_2 \right \|^2}{\sigma^2} \right )}.
\end{equation}

The outage and the ergodic performance of the SR with $K=1$ are analyzed in \cite{zhang2019backscatter} and the analytical results show that at high SNR region, the ergodic rate of the secondary transmission increases by about $3$ bit/s/Hz when the SNR increases $10$ dB. For large $K$, the ergodic rate of SR under multi-antenna PTx and PRx are analyzed in \cite{zhou2019ergodic} and the upper bound of the secondary rate increases with the number of receive antennas $M_r$ and decreases with the transmission period $K$, scaling like $\frac{1}{K}\log_2(KM_r)$ at high SNR region. The achievable rate region of SR for large $K$ under binary modulation at STx is analyzed in \cite{liu2018backscatter} and the results exhibit that the rate region of SRN is strictly larger than that of the conventional \emph{time division multiple access} (TDMA) scheme.

\subsection{Single-Input-Single-Output (SISO) Primary Channel}

For \emph{single-input-single-output} (SISO) primary channel, the PTx is able to control its transmit power and the STx is able to adjust its reflection efficiency to fulfill specific performance metric requirements. The power and reflection efficiency constraints can be described as follows:
\begin{itemize}
\item \emph{Peak Transmit Power Constraint}: Let $P_{pk}$ be the peak power available at the transmitter, the power constraint is given by
\begin{equation}\label{eq:peak}
p\leq P_{pk}.
\end{equation}

\item \emph{Average Transmit Power Constraint}: When the long-term power budget $P_{av}$ is considered, we have
\begin{equation}\label{eq:average}
\bbE[p]\leq P_{av}.
\end{equation}
The expectation is usually taken over the channel realizations.

\item \emph{Reflection Efficiency Constraint}: The reflection efficiency $\alpha$ can be adjusted in each fading block. It needs to satisfy the constraint:
\begin{equation}\label{eq:Mismatch}
  0\leq\alpha\leq1.
\end{equation}
\end{itemize}

%
With the aforementioned constraints, the weighted sum ergodic rate maximization problem under the peak/average power constraint is considered in \cite{guo2019resource}. Specifically, by jointly optimizing the transmit power $p$ at the PTx and the reflection efficiency $\alpha$ at the STx, the ergodic weighted sum rate is maximized under either long-term or short-term transmit-power constraint over the fading states. It is shown that the larger the $\alpha$ is, the better performance for both systems is obtained, due to the mutualism of the two systems.
Compared to the ergodic rate optimization in the CR networks \cite{Kang2009Optimal} where the secondary system has to keep its interference level to the primary receiver below a pre-designed threshold, the STx in the SR does not need to consider the interference constraint.
The power allocation problem under the average power constraint for three SR paradigms is considered in \cite{guo2019cooperative} and the outage probabilities for these  paradigms based on the optimized $\alpha$ and $p$ are analyzed in \cite{ding2019outage}. It is shown that under different transmission schemes, there are different relationships between the achievable rates of the primary and secondary transmissions.

\subsection{Multiple-Input-Single-Output (MISO) Primary Channel}

For \emph{multiple-input-single-output} (MISO) primary channel, spatial degrees of freedom can be exploited to balance the objectives of the primary and secondary systems. Here are some constraints to be considered:
\begin{itemize}

\item \emph{Transmit Beamforming Constraint}: For a given power budget $P_t$, the transmit beamforming vector ${\bv}$ should satisfy the following constraint
\begin{equation}
\left\|\mathbf{v}\right\|^2\leq P_t.\label{eq:Beamforming_Constraint}
\end{equation}
\item \emph{Primary Transmission Rate Constraint}:
In the scenario where the primary system is sensitive to its transmit rate, the rate constraint should be considered, which is described as
\begin{equation}
R_{s}\geq \tilde{R}_s,\label{eq:cellular_rate}
\end{equation}
where $\tilde{R}_s$ is the minimum primary transmission rate requirement.
\item \emph{Secondary Transmission Rate Constraint}:
The secondary transmission rate constraint is given by
\begin{equation}
R_{c}\geq \tilde{R}_c,\label{eq:MTC_SNR}
\end{equation}
where $\tilde{R}_c$ is the minimum secondary transmission rate requirement.
\end{itemize}

Considering the above constraints, by optimizing the transmit beamforming vector and the transmit power at the PTx, the weighted sum rate maximization problem and the transmit power minimization problem have been studied in \cite{long2019symbiotic}.
By solving these optimization problems, the proposed SR not only enables the opportunistic transmission for the secondary system, but also enhances the primary transmission by properly exploiting the additional signal path provided by the secondary transmission.
Moreover, the achievable rate of STx in finite block-length regime is considered in \cite{chu2020resource} and both of transmit power minimization and energy efficiency maximization optimization problems under the achievable throughput constraints are formulated to design the beamforming vector at the PTx.

\subsection{Full-duplex STx}

The STx can be designed with full-duplex function \cite{liu2017full,long2019full}, where the STx is able to absorb a fraction of the incident signal to decode the messages from PTx, and simultaneously transmit its own information to the SRx by backscattering the remaining part of the incident signal.
In \cite{long2019full}, a full-duplex STx is considered in the SR and the achievable rates of secondary system with Gaussian and QAM codewords are derived, based on which, the power minimization problem is formulated to design the reflection efficiency at the STx as well as the beamforming vector at the PTx. The results show that the SR with full-duplex STx outperforms the optimally designed half-duplex scheme.


\section{RIS-assisted SR (RSR)}\label{sec:RIS}


RIS is a two-dimensional artificial structure including multiple reflecting elements and each reflecting element is able to reflect the incident signal with different reflection coefficients. With the assistance of RIS, the SR system can capture the direction of the signal from the PRx and enhance the strength of the backscattering link.
For RSR, the STx in Fig.~\ref{fig:system}(a) is just an RIS. The PTx transmits information to the PRx, and the RIS as a STx embeds its messages over the primary RF signals to the SRx by using backscattering radio technology.

In RSR, let $M_b$ be the number of reflecting elements at RIS, $s(t)$ the baseband form signal transmitted from PTx, $\theta_{m}(t)$ the reflection coefficient at the $m$th element, the backscattering link signal received at SRx in baseband form becomes
\begin{equation}
    y(t)=\sum_{m=1}^{M_b}g_{m} \theta_{m} (t) l_{m}s (t)+u(t),
    \label{eq:backscattering_link}
\end{equation}
where $l_m$ is the channel response from the PTx to the \emph{m}th element of RIS, $g_{m}$ is the one from the \emph{m}th element to the SRx, and $u(t)$ is the additive noise at the receiver.
Generally, the reflecting elements are passive and both the amplitudes and phase shifts of each element can be adjusted, yielding $|\theta_{m}(t)|^2 \leq 1,\forall m$. For some applications, constant amplitude constraint is imposed on $\theta_{m}(t)$ and in this case, only the phase shifts of the reflecting elements can be adjusted. Thus when the STx does not carry its own massages, the reflection coefficients can be chosen such that the objectives of the primary system are maximized. Thus the actively studied RIS \cite{liang2019large,Wu2019,Huang2019,Guo2020,zhao2019survey,basar2020reconfigurable,kammoun2020asymptotic,di2019smart,jamali2019intelligent,bjornson2019intelligent,jung2020performance} is in fact a special case of RSR.

When the reflection coefficients, $\theta_{m}(t)$, are time-invariant and are chosen properly, the received SNR in \eqref{eq:backscattering_link} of the backscattering link increases quadratically with the number of reflecting elements \cite{liang2019large,Wu2019,Huang2019,Guo2020}.
When backscattering modulation is used, the information at the RIS can be represented by the time-varying reflection coefficients, through which beneficial performance gain to the primary transmission has also been observed \cite{zhang2020sr, zhang2020large}. With reference to Fig.~\ref{fig:system}(a), even when the direct link from PTx-PRx is considered, the backscattering link from PTx-RIS-PRx can still enhance the performance of primary transmission.
Thus, RSR is one solution to enhance the backscattering link signal to further enhance the performance of the primary and secondary transmissions.


On the other hand, as long as the RIS has the channel information, it can carry out passive beamforming by designing the reflection coefficients $\theta_{m}, \forall m$, with which, the RSR system can capture the intended PTx signal instead of backscattering all the ambient signals, and thus it can avoid undesired interference at the SRx \cite{zhang2020large}.

In \cite{zhang2020sr} and \cite{zhang2020large}, RSR is considered to enable the secondary transmission as well as to enhance the primary transmission via intelligently reconfiguring the wireless communications. By jointly designing the active transmit beamforming at the PTx and the backscattering beamforming at the RIS, the achievable rate for the secondary system is maximized with MISO primary channel under the primary rate constraint and other practical constraint in \cite{zhang2020sr}, and the transmit power is minimized with MIMO primary channel under some rate constraints for the SR in \cite{zhang2020large}. Both of the above studies show that with proper beamforming design, not only the secondary transmission can be successfully conducted, but also the primary performance with RIS assistance is superior to that without RIS.

\section{Full-Duplex SR (FDSR)}\label{sec:full}

\begin{figure}
\centering
\includegraphics[width=0.8\columnwidth] {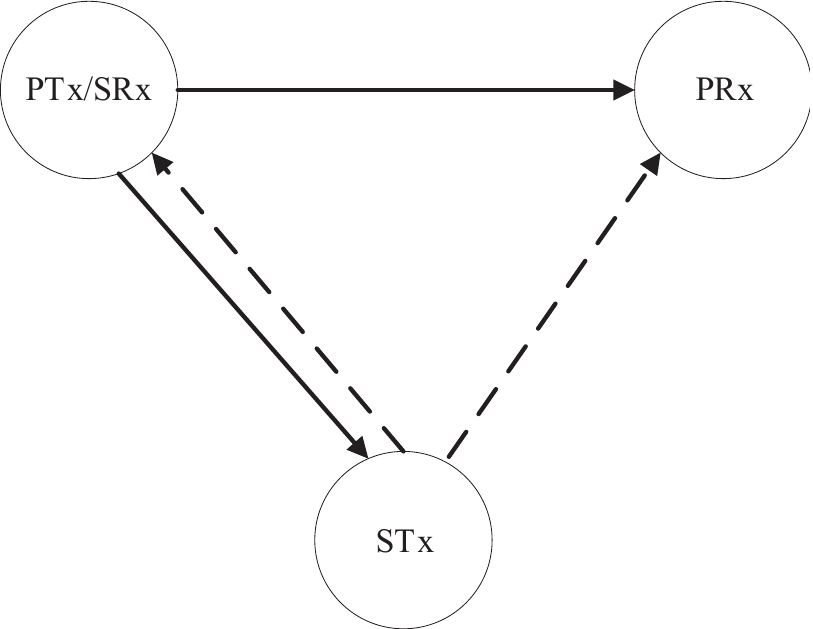}
\caption{System model for full-duplex SR: the PTx transmits primary signal to the PRx, and receives the STx signals as a SRx. Solid lines denote transmissions from active radios, while dash lines denote reflections from backscattering radios.}
\label{fig:system6}
\end{figure}

If the PTx is with full-duplex function, as shown in Fig.~\ref{fig:system6}, the PTx can not only transmit primary signal to the PRx but also receive the STx signals as a SRx, which is called FDSR system.

Denote by $\beta_1$ the self-interference channel at the PTx and $\beta_2$ the composite channel coefficient from the PTx to the STx and further to the PTx. The received signal at the PTx can be written as
\begin{align}
y_k(n) = {\sqrt{p} \beta_1 s_k(n)}+ {\sqrt{p} \beta_2 s_k(n) c(n)} + u_k(n), \label{eq:received5}
\end{align}
where $u_{k}(n) \sim \mathcal{CN}(0,\sigma{^2})$. Clearly, when the PTx recovers the STx message, it suffers from the self-interference caused by the transmitted signal. Due to the fact that the PTx has the exact information of $s_k(n)$, the first term in \eqref{eq:received5} can be canceled by using the self-interference cancellation technique. As such, the STx messages can be recovered with a high performance.

The FDSR system is firstly proposed in \cite{bharadia2015backfi}, where the WiFi AP is designed to have the full-duplex function and the WiFi AP can decode the backscatter signals from the STx in the process of transmitting data to normal WiFi users. The prototypes and experiments show that the communication rate can achieve $5$ Mbps at a range of $1$ m.
The information-theoretic capacity of both the primary and the secondary systems of FDSR is derived in \cite{darsena2017modeling}. The analytical results show that the primary system can turn the backscatter signal into a form of multipath diversity under some reasonable conditions and the secondary transmission can achieve significant data rate with relatively short distances.
In \cite{smida2017reflectfx}, throughput and ergodic capacity expressions are derived for the FDSR system with a full-duplex STx and the simulation results show that this FDSR system outperforms both conventional full-duplex and half-duplex.
Furthermore, resource allocation at the relevant nodes are conducted to fulfill the desired performance requirements.
In \cite{yang2018optimal}, the PTx transmits OFDM signals to the PRx and simultaneously receives the signals backscattered from multiple STxs in a TDMA manner. The backscatter time and reflection efficiency of each STx, and the subcarrier power allocation of the PTx are designed by maximizing the minimum throughput among all STxs under the primary rate constraint and other practical constraints. In addition, the single STx case with closed-form solutions is considered and the throughput region, which characterizes the Pareto-optimal throughput trade-offs among all STxs, is analyzed.
In \cite{xiao2019resource}, the PTx transmits information to multiple PRxs by using TDMA and multiple STxs backscatter messages to the PTx by using TDMA.
The sum secondary transmission rate maximization problem is studied to design the backscatter time, transmit power at the PTx, and the reflection efficiency at each STx under the primary rate constraint.

\section{Applications} \label{sec:application}



The mutualism spectrum sharing and low power consumption nature of SR makes it suitable as a massive access technology for 6G and beyond \cite{nawaz2020non,bariah2020prospective,chen2020vision,8820755}. In this section, we address three emerging applications of SR, including e-health, smart home, and environmental monitoring, riding over different types of primary networks.

\subsection{E-health}
Human health monitoring is one of the most promising applications for 6G. By placing a number of health monitoring sensors, such as pulse and temperature monitors, implantable devices on the body, instant health status can be transmitted to a central unit \cite{otto2006system}. The size and power consumption for health monitoring devices are two major concerns for system design\cite{report2020from}. Traditional wireless sensor communication technologies, such as Bluetooth and Zigbee, are not suitable for this application since the active communication scheme will cause high power consumption and the heat generated during communications may be harmful to the body, especially for the implantable sensors.

SR is a promising technology to overcome the above challenges due to the following reasons. First, the information transmission uses a passive way and thus the power consumption is ultra-low. Second, the basic backscattering radio circuits is very simple, so the size of the health monitoring devices can be very tiny.
Last but not least, the health monitoring sensors can use the existing cellular infrastructure to transmit messages, which is beneficial and economical for
service providers. For example, the user's mobile phone can act as the center for health information collection by using the uplink/downlink cellular signals.


\subsection{Smart Home}
Smart home is another promising application in future wireless networks, in which almost all ubiquitous electronic devices within a house are connected to build an automatic and intelligent home environment \cite{ricquebourg2006smart}.
In such a scenario, communications among different electronic devices generally feature low data rate and low power consumption, which places SR as the promising solution for smart home by backscattering the WiFi signals in the house. Specifically, a WiFi \emph{access point} (AP) installed in the house or the user's mobile phone can be the information center for monitoring the home environment status, such as air quality, room temperature, and usage of electronic devices.

\subsection{Environmental Monitoring}

Environmental monitoring represents the processes for continuously monitoring the environment status, such as atmospheric and weather conditions, through deploying wireless sensors \cite{messer2006environmental}. The power consumption for the wireless sensors and the transmission range are two major concerns for the system design. SR can be considered as an environmental monitoring network due to its low cost and low power-consumption. Specifically, all the monitoring devices sense the environmental conditions and then feedback them to the receivers by modulating their messages over the strong broadcast TV signals. 
Compared with the other wireless techniques, such as LoRa and NB-IoT, the SR technology uses the existing cellular network infrastructure with lower power consumption and low cost, and thus is easier to be widely deployed.


\section{Challenges and Opportunities} \label{sec:chall}

SR is a promising solution to achieve spectrum-, power- and cost-efficient communications. There are however some critical limitations, challenges, and open research problems, which need to be addressed.




\subsection{Channel Modeling and Estimation}

It is important to establish accurate channel models to capture the essential behavior of the backscattering channels. In particular, due to the multiplicative nature of the backscattering channels, there are additional channel properties to be considered, such as rank-deficient, channel correlation and so on. There are some channel estimation schemes to estimate the cascade backscattering channels, e.g.,\cite{he2019cascaded,chen2019channel,taha2019enabling,mirza2019channel}. However, most of them are with high complexity. It is highly desired to design low complexity algorithm or conduct joint channel estimation and signal detection.
Besides, considering the massive devices, pilot symbols need to be carefully designed to support effective channel estimation, yet to avoid pilot contamination.

\subsection{Hardware Design}

The synchronization between the primary and secondary symbols at the STx is a considerable problem. The carrier phase and timing recovery circuitry in the traditional synchronization algorithm requires the oscillator component, which is power-consuming \cite{goldsmith2005wireless}.
Thus, it is very important to design a low complexity signal synchronization algorithm for the STx to track the primary signals on the RF level .
Another considerable problem at the receiver is how to capture the weak backscattering link signal from the strong direct link signal.
The use of active load or multiple antennas at the STx is one solution to enhance the strength of the backscattering link signal. However, considering the fact that both active load and the use of multiple antennas consume power, it remains an open problem on how to design energy-efficient active load and multiple antenna solutions.



\subsection{Security and Privacy}

Security and privacy are always critical in wireless communication systems. Due to the symbiotic relationship between the primary and secondary transmissions in SR, if one attacker disrupts the primary transmissions, the secondary transmission may be affected. It is an urgent problem to design a suitable and effective policy to ensure the security and privacy of the SR.

\section {Conclusions}

In this paper, we have provided a systematic view on SR, a cognitive backscattering communication system, and have highlighted the large potential of SR in future wireless networks to achieve: (1) enhanced spectrum efficiency using mutualism spectrum sharing; and (2) enhanced energy efficiency through highly reliable backscattering communications. Three fundamental aspects of SR are discussed, including system model, transceiver design, and resource allocation schemes. RIS-assisted SR and full-duplex SR are also presented. Emerging applications and some design challenges are discussed. SR is a communication system with hybrid active and backscattering radios, and the actively studied RIS is a special case of the SR proposed in this paper. We foresee that the future wireless networks will be having more hybrid radios in the system \cite{Larsson-weave}, and hope this article provide an effective guidance for the future work in this emerging and fantastic field.



\bibliographystyle{IEEEtran}

\begin{thebibliography}{10}
\providecommand{\url}[1]{#1}
\csname url@samestyle\endcsname
\providecommand{\newblock}{\relax}
\providecommand{\bibinfo}[2]{#2}
\providecommand{\BIBentrySTDinterwordspacing}{\spaceskip=0pt\relax}
\providecommand{\BIBentryALTinterwordstretchfactor}{4}
\providecommand{\BIBentryALTinterwordspacing}{\spaceskip=\fontdimen2\font plus
\BIBentryALTinterwordstretchfactor\fontdimen3\font minus
  \fontdimen4\font\relax}
\providecommand{\BIBforeignlanguage}[2]{{%
\expandafter\ifx\csname l@#1\endcsname\relax
\typeout{** WARNING: IEEEtran.bst: No hyphenation pattern has been}%
\typeout{** loaded for the language `#1'. Using the pattern for}%
\typeout{** the default language instead.}%
\else
\language=\csname l@#1\endcsname
\fi
#2}}
\providecommand{\BIBdecl}{\relax}
\BIBdecl

\bibitem{imt2015}
``{IMT} traffic estimates for the years 2020 to 2030,'' Available:
  https://www.itu.int.

\bibitem{matti2019key}
M.~Latva-aho and K.~Leppanen, ``Key drivers and research challenges for 6{G}
  ubiquitous wireless intelligence,'' \emph{University of Oulu, While Paper},
  2019. Available: http://urn.fi/urn:isbn:9789526223544.

\bibitem{e2016Identification}
E.~Union, ``Identification and quantification of key socio-economic data to
  support strategic planning for the introduction of {5G in Europe},'' 2016.

\bibitem{mitola1999cognitive}
J.~Mitola~III and G.~Q. Maguire~Jr, ``Cognitive radio: making software radios
  more personal,'' \emph{IEEE Pers. Commun.}, vol.~6, no.~4, pp. 13--18, 1999.

\bibitem{haykin2005cognitive}
S.~Haykin, ``Cognitive radio: brain-empowered wireless communications,''
  \emph{IEEE J. Sel. Areas Commun.}, vol.~23, no.~2, pp. 201--220, 2005.

\bibitem{liang2011cognitive}
Y.-C. Liang, K.-C. Chen, G.~Y. Li, and P.~Mahonen, ``Cognitive radio networking
  and communications: An overview,'' \emph{IEEE Trans. Veh. Tech.}, vol.~60,
  no.~7, pp. 3386--3407, 2011.

\bibitem{qin201920}
Z.~{Qin}, X.~{Zhou}, L.~{Zhang}, Y.~{Gao}, Y.-C. Liang, and G.~Y. {Li}, ``20
  years of evolution from cognitive to intelligent communications,'' \emph{IEEE
  Trans. Cogn. Commun. Netw.}, vol.~6, no.~1, pp. 6--20, 2020.

\bibitem{report2020from}
``From healthcare to homecare,'' \emph{{Report of ericsson.com}}, 2020.
  {Available:
  https://www.ericsson.com/en/reports-and-papers/consumerlab/reports/transforming-healthcare-homecare}.

\bibitem{liu2013ambient}
V.~Liu, A.~Parks, V.~Talla, S.~Gollakota, D.~Wetherall, and J.~R. Smith,
  ``Ambient backscatter: Wireless communication out of thin air,'' in
  \emph{Proc. of ACM SIGCOMM}, vol.~43, no.~4.\hskip 1em plus 0.5em minus
  0.4em\relax Hong Kong, China: ACM, Aug. 2013, pp. 39--50.

\bibitem{Boyer2014Backscatter}
C.~Boyer and S.~Roy, ``Backscatter communication and {RFID}: {Coding}, energy,
  and {MIMO} analysis,'' \emph{IEEE Trans. Commun.}, vol.~62, no.~3, pp.
  770--785, Mar. 2014.

\bibitem{van2018ambient}
N.~Van~Huynh, D.~T. Hoang, X.~Lu, D.~Niyato, P.~Wang, and D.~I. Kim, ``Ambient
  backscatter communications: A contemporary survey,'' \emph{IEEE Commun.
  Surveys Tuts.}, vol.~20, no.~4, pp. 2889--2922, 2018.

\bibitem{wang2016ambient}
G.~Wang, F.~Gao, R.~Fan, and C.~Tellambura, ``Ambient backscatter communication
  systems: Detection and performance analysis,'' \emph{IEEE Trans. Commun.},
  vol.~64, no.~11, pp. 4836--4846, Nov. 2016.

\bibitem{qian2017noncoherent}
J.~Qian, F.~Gao, G.~Wang, S.~Jin, and H.~Zhu, ``Noncoherent detections for
  ambient backscatter system,'' \emph{IEEE Trans. Wireless Commun.}, vol.~16,
  no.~3, pp. 1412--1422, Mar. 2017.

\bibitem{qian2017semi}
------, ``Semi-coherent detection and performance analysis for ambient
  backscatter system,'' \emph{IEEE Trans. Commun.}, vol.~65, no.~12, pp.
  5266--5279, Dec. 2017.

\bibitem{qian2018iot}
J.~Qian, A.~N. Parks, J.~R. Smith, F.~Gao, and S.~Jin, ``{IoT} communications
  with $ {M }$-{PSK} modulated ambient backscatter: Algorithm, analysis, and
  implementation,'' \emph{IEEE Internet Things J.}, vol.~6, no.~1, pp.
  844--855, 2018.

\bibitem{liu2017coding}
Y.~Liu, G.~Wang, Z.~Dou, and Z.~Zhong, ``Coding and detection schemes for
  ambient backscatter communication systems,'' \emph{IEEE Access}, vol.~5, pp.
  4947--4953, 2017.

\bibitem{Iyery_AmBC-frquency-fisf2016}
V.~{Iyery}, V.~{Tallay}, B.~{Kelloggy}, S.~{Gollakota}, and J.~R. Smith,
  ``Inter-technology backscatter: Towards internet connectivity for implanted
  devices,'' in \emph{2016 Proc. of ACM SIGCOMM}.\hskip 1em plus 0.5em minus
  0.4em\relax Florianopolis, Brazil: ACM, August 2016, pp. 356--369.

\bibitem{zhang_AmBC-frquency-fisf2016}
P.~{Zhang}, M.~{Rostami}, P.~{Hu}, and D.~Ganesan, ``Enabling practical
  backscatter communication for on-body sensors,'' in \emph{2016 Proc. of ACM
  SIGCOMM}.\hskip 1em plus 0.5em minus 0.4em\relax Florianopolis, Brazil: ACM,
  August 2016, pp. 370--383.

\bibitem{elmossallamy2019noncoherent}
M.~A. ElMossallamy, M.~Pan, R.~J{\"a}ntti, K.~G. Seddik, G.~Y. Li, and Z.~Han,
  ``Noncoherent backscatter communications over ambient {OFDM} signals,''
  \emph{IEEE Trans. Commun.}, vol.~67, no.~5, pp. 3597--3611, 2019.

\bibitem{yang2016backscatter}
G.~Yang and Y.-C. Liang, ``Backscatter communications over ambient {OFDM}
  signals: Transceiver design and performance analysis,'' in \emph{Proc. of
  IEEE Glob. Commun. Conf. (GLOBECOM)}.\hskip 1em plus 0.5em minus 0.4em\relax
  Washington, D.C.: IEEE, Dec. 2016, pp. 1--6.

\bibitem{yang2018modulation}
G.~Yang, Y.-C. Liang, R.~Zhang, and Y.~Pei, ``Modulation in the air:
  Backscatter communication over ambient {OFDM} carrier,'' \emph{IEEE Trans.
  Commun.}, vol.~66, no.~3, pp. 1219--1233, Mar. 2018.

\bibitem{nguyen2019signal}
T.~L. Nguyen, Y.~Shin, J.~Y. Kim, and D.~I. Kim, ``Signal detection for ambient
  backscatter communication with {OFDM} carriers,'' \emph{Sensors}, vol.~19,
  no.~3, p. 517, 2019.

\bibitem{zhang2018interference}
C.~Zhang, G.~Wang, P.~D. Diamantoulakis, F.~Gao, and G.~K. Karagiannidis,
  ``Interference-free transceiver design and signal detection for ambient
  backscatter communication systems over frequency-selective channels,''
  \emph{arXiv preprint arXiv:1812.11278}, 2018.

\bibitem{YangLiangZhangICC17}
G.~Yang, Y.-C. Liang, and Q.~Zhang, ``Cooperative receiver for ambient
  backscatter communications with multiple antennas,'' in \emph{Proc. of {IEEE}
  Conf. Commun. (ICC)}.\hskip 1em plus 0.5em minus 0.4em\relax Paris: IEEE, May
  2017, pp. 1--6.

\bibitem{yang2018cooperative}
G.~Yang, Q.~Zhang, and Y.-C. Liang, ``Cooperative ambient backscatter
  communications for green {Internet-of-Things},'' \emph{IEEE Internet Things
  J.}, vol.~5, no.~2, pp. 1116--1130, Apr. 2018.

\bibitem{liu2018backscatter}
W.~Liu, Y.-C. Liang, Y.~Li, and B.~Vucetic, ``Backscatter multiplicative
  multiple-access systems: Fundamental limits and practical design,''
  \emph{IEEE Trans. Wireless Commun.}, vol.~17, no.~9, pp. 5713--5728, 2018.

\bibitem{guo2019cooperative}
H.~Guo, Y.-C. Liang, R.~Long, and Q.~Zhang, ``Cooperative ambient backscatter
  system: A symbiotic radio paradigm for passive {IoT},'' \emph{IEEE Wireless
  Commun Lett.}, vol.~8, no.~4, pp. 1191--1194, 2019.

\bibitem{zhang2019backscatter}
Q.~{Zhang}, L.~{Zhang}, Y.-C. Liang, and P.~{Kam}, ``Backscatter-{NOMA}: A
  symbiotic system of cellular and {Internet-of-Things} networks,'' \emph{IEEE
  Access}, vol.~7, pp. 20\,000--20\,013, 2019.

\bibitem{long2019full}
R.~{Long}, H.~{Guo}, L.~{Zhang}, and {Y.-C. Liang}, ``Full-duplex backscatter
  communications in symbiotic radio systems,'' \emph{IEEE Access}, vol.~7, pp.
  21\,597--21\,608, 2019.

\bibitem{guo2019resource}
H.~Guo, Y.-C. Liang, R.~Long, S.~Xiao, and Q.~Zhang, ``Resource allocation for
  symbiotic radio system with fading channels,'' \emph{IEEE Access}, vol.~7,
  pp. 34\,333--34\,347, 2019.

\bibitem{long2019symbiotic}
R.~Long, Y.-C. Liang, H.~Guo, G.~Yang, and R.~Zhang, ``Symbiotic radio: A new
  communication paradigm for passive internet-of-things,'' \emph{IEEE Internet
  Things J.}, vol.~7, pp. 1350--1363, 2020.

\bibitem{nawaz2020non}
S.~J. Nawaz, S.~K. Sharma, B.~Mansoor, M.~N. Patwary, and N.~M. Khan,
  ``Non-coherent and backscatter communications: Enabling ultra-massive
  connectivity in {6G} wireless networks,'' \emph{arXiv preprint
  arXiv:2005.10937}, 2020.

\bibitem{bariah2020prospective}
L.~Bariah, L.~Mohjazi, S.~Muhaidat, P.~C. Sofotasios, G.~K. Kurt,
  H.~Yanikomeroglu, and O.~A. Dobre, ``A prospective look: Key enabling
  technologies, applications and open research topics in {6G} networks,''
  \emph{arXiv preprint arXiv:2004.06049}, 2020.

\bibitem{chen2020vision}
S.~Chen, Y.-C. Liang, S.~Sun, S.~Kang, W.~Cheng, and M.~Peng, ``Vision,
  requirements, and technology trend of {6G}: how to tackle the challenges of
  system coverage, capacity, user data-rate and movement speed,'' \emph{IEEE
  Wireless Commun.}, vol.~27, no.~2, pp. 218--228, 2020.

\bibitem{8820755}
L.~{Zhang}, Y.-C. Liang, and D.~{Niyato}, ``{6G} visions: Mobile
  ultra-broadband, super internet-of-things, and artificial intelligence,''
  \emph{China Commun.}, vol.~16, no.~8, pp. 1--14, 2019.

\bibitem{Hansen1989}
R.~C. {Hansen}, ``Relationships between antennas as scatterers and as
  radiators,'' \emph{Proc. the IEEE}, vol.~77, no.~5, pp. 659--662, 1989.

\bibitem{Thomas2012}
S.~J. {Thomas} and M.~S. {Reynolds}, ``A 96 {Mbit/sec}, 15.5 pj/bit 16-{QAM}
  modulator for {UHF} backscatter communication,'' in \emph{Proc. of IEEE Int.
  Conf. RFID (RFID)}.\hskip 1em plus 0.5em minus 0.4em\relax Orlando, FL, USA:
  IEEE, Apr. 2012, pp. 185--190.

\bibitem{Thomas2012Quadrature}
S.~J. Thomas, E.~Wheeler, J.~Teizer, and M.~S. Reynolds, ``Quadrature amplitude
  modulated backscatter in passive and semipassive uhf rfid systems,''
  \emph{IEEE Trans. Microw. Theory Techn.}, vol.~60, no.~4, pp. 1175--1182,
  April 2012.

\bibitem{Amato2018}
F.~{Amato}, C.~W. {Peterson}, B.~P. {Degnan}, and G.~D. {Durgin}, ``Tunneling
  {RFID} tags for long-range and low-power microwave applications,'' \emph{IEEE
  J. Radio Freq. Identif.}, vol.~2, no.~2, pp. 93--103, Jun. 2018.

\bibitem{1965APL}
G.~T. Munsterman, ``Tunnel-diode microwave amplifiers,'' \emph{APL Tech. Dig.},
  pp. 2--10, 1965.

\bibitem{Amato2018a}
F.~{Amato}, H.~M. {Torun}, and G.~D. {Durgin}, ``{RFID} backscattering in
  long-range scenarios,'' \emph{IEEE Trans. Wireless Commun.}, vol.~17, no.~4,
  pp. 2718--2725, April 2018.

\bibitem{Khaledian2019}
S.~{Khaledian}, F.~{Farzami}, H.~{Soury}, B.~{Smida}, and D.~{Erricolo},
  ``Active two-way backscatter modulation: An analytical study,'' \emph{IEEE
  Trans. Wireless Commun.}, vol.~18, no.~3, pp. 1874--1886, Mar. 2019.

\bibitem{Bousquet20124}
J.~{Bousquet}, S.~{Magierowski}, and G.~G. {Messier}, ``A {4-GHz} active
  scatterer in 130-nm {CMOS} for phase sweep amplify-and-forward,'' \emph{IEEE
  Trans. Circuits Syst. I, Reg. Papers}, vol.~59, no.~3, pp. 529--540, Mar.
  2012.

\bibitem{guo2019cognitive}
H.~Guo, R.~Long, and Y.-C. Liang, ``Cognitive backscatter network: A spectrum
  sharing paradigm for passive {IoT},'' \emph{IEEE Wireless Commun. Lett.},
  vol.~8, no.~5, pp. 1423--1426, 2019.

\bibitem{ruttik2018does}
K.~Ruttik, R.~Duan, R.~J{\"a}ntti, and Z.~Han, ``Does ambient backscatter
  communication need additional regulations?'' in \emph{Proc. of IEEE
  DySPAN}.\hskip 1em plus 0.5em minus 0.4em\relax Seoul, South Korea: IEEE,
  Oct. 2018, pp. 1--6.

\bibitem{zhang2019constellation}
Q.~Zhang, H.~Guo, Y.-C. Liang, and X.~Yuan, ``Constellation learning-based
  signal detection for ambient backscatter communication systems,'' \emph{IEEE
  J. Sel. Areas Commun.}, vol.~37, no.~2, pp. 452--463, 2019.

\bibitem{tian2018modulation}
J.~Tian, Y.~Pei, Y.-D. Huang, and Y.-C. Liang, ``Modulation-constrained
  clustering approach to blind modulation classification for {MIMO} systems,''
  \emph{IEEE Trans. Cogn. Commun. Netw.}, vol.~4, no.~4, pp. 894--907, 2018.

\bibitem{wang2019machine}
X.~Wang, R.~Duan, H.~Yigitler, E.~Menta, and R.~Jantti, ``Machine
  learning-assisted detection for {BPSK}-modulated ambient backscatter
  communication systems,'' in \emph{Proc. of IEEE Glob. Commun. Conf.
  (GLOBECOM)}.\hskip 1em plus 0.5em minus 0.4em\relax Waikoloa, HI, USA: IEEE,
  Dec. 2019, pp. 1--6.

\bibitem{aborahama2019stochastic}
Y.~Aborahama and M.~S. Hassan, ``On the stochastic modeling of the holding time
  of {SUs to PU} channels in cognitive radio networks,'' \emph{IEEE Trans.
  Cogn. Commun. Netw.}, vol.~6, no.~1, pp. 282--295, 2019.

\bibitem{huang2019achieving}
Y.~Huang, Y.~Chen, Y.~T. Hou, and W.~Lou, ``Achieving fair {LTE/Wi-Fi}
  coexistence with real-time scheduling,'' \emph{IEEE Trans. Cogn. Commun.
  Netw.}, vol.~6, no.~1, pp. 366--380, 2019.

\bibitem{ghasemi2007fundamental}
A.~Ghasemi and E.~S. Sousa, ``Fundamental limits of spectrum-sharing in fading
  environments,'' \emph{IEEE Trans. Wireless Commun.}, vol.~6, no.~2, pp.
  649--658, 2007.

\bibitem{musavian2009capacity}
L.~Musavian and S.~A{\"\i}ssa, ``Capacity and power allocation for
  spectrum-sharing communications in fading channels,'' \emph{IEEE Trans.
  Wireless Commun.}, vol.~8, no.~1, pp. 148--156, 2009.

\bibitem{Kang2009Optimal}
X.~{Kang}, Y.-C. {Liang}, A.~{Nallanathan}, H.~K. {Garg}, and R.~{Zhang},
  ``Optimal power allocation for fading channels in cognitive radio networks:
  Ergodic capacity and outage capacity,'' \emph{IEEE Trans. Wireless Commun.},
  vol.~8, no.~2, pp. 940--950, 2009.

\bibitem{tao2018symbol}
Q.~Tao, C.~Zhong, H.~Lin, and Z.~Zhang, ``Symbol detection of ambient
  backscatter systems with manchester coding,'' \emph{IEEE Trans. Wireless
  Commun.}, vol.~17, no.~6, pp. 4028--4038, 2018.

\bibitem{zhou2019ergodic}
S.~Zhou, W.~Xu, K.~Wang, C.~Pan, M.-S. Alouini, and A.~Nallanathan, ``Ergodic
  rate analysis of cooperative ambient backscatter communication,'' \emph{IEEE
  Wireless Commun. Lett.}, vol.~8, no.~6, pp. 1679--1682, 2019.

\bibitem{ding2019outage}
H.~Ding, D.~B. da~Costa, and J.~Ge, ``Outage analysis for cooperative ambient
  backscatter systems,'' \emph{IEEE Wireless Commun. Lett.}, vol.~9, no.~5, pp.
  601--605, 2019.

\bibitem{chu2020resource}
Z.~Chu, W.~Hao, P.~Xiao, M.~Khalily, and R.~Tafazolli, ``Resource allocations
  for symbiotic radio with finite block length backscatter link,'' \emph{IEEE
  Internet Things J.}, DOI: 10.1109/JIOT.2020.2980928, 2020.

\bibitem{liu2017full}
W.~Liu, K.~Huang, X.~Zhou, and S.~Durrani, ``Full-duplex backscatter
  interference networks based on time-hopping spread spectrum,'' \emph{IEEE
  Trans. Wireless Commun.}, vol.~16, no.~7, pp. 4361--4377, 2017.

\bibitem{liang2019large}
Y.-C. Liang, R.~Long, Q.~Zhang, J.~Chen, H.~V. Cheng, and H.~Guo, ``Large
  intelligent surface/antennas {(LISA)}: Making reflective radios smart,''
  \emph{J. Commun. Inf. Netw.}, vol.~4, no.~2, pp. 40--50, 2019, also available
  at arXiv:1906.06578.

\bibitem{Wu2019}
Q.~{Wu} and R.~{Zhang}, ``Intelligent reflecting surface enhanced wireless
  network via joint active and passive beamforming,'' \emph{IEEE Trans.
  Wireless Commun.}, vol.~18, no.~11, pp. 5394--5409, Nov 2019.

\bibitem{Huang2019}
C.~{Huang}, A.~{Zappone}, G.~C. {Alexandropoulos}, M.~{Debbah}, and C.~{Yuen},
  ``Reconfigurable intelligent surfaces for energy efficiency in wireless
  communication,'' \emph{IEEE Trans. Wireless Commun.}, vol.~18, no.~8, pp.
  4157--4170, Aug 2019.

\bibitem{Guo2020}
H.~{Guo}, Y.-C. {Liang}, J.~{Chen}, and E.~G. {Larsson}, ``Weighted sum-rate
  maximization for reconfigurable intelligent surface aided wireless
  networks,'' \emph{IEEE Trans. Wireless Commun.}, vol.~19, no.~5, pp.
  3064--3076, 2020.

\bibitem{zhao2019survey}
J.~Zhao and Y.~Liu, ``A survey of intelligent reflecting surfaces {(IRSs):
  Towards 6G} wireless communication networks,'' \emph{arXiv preprint
  arXiv:1907.04789}, 2019.

\bibitem{basar2020reconfigurable}
E.~Basar, ``Reconfigurable intelligent surface-based index modulation: A new
  beyond {MIMO paradigm for 6G},'' \emph{IEEE Trans. Commun.}, vol.~68, no.~5,
  pp. 3187--3196, 2020.

\bibitem{kammoun2020asymptotic}
A.~Kammoun, A.~Chaaban, M.~Debbah, M.-S. Alouini \emph{et~al.}, ``Asymptotic
  max-min {SINR} analysis of reconfigurable intelligent surface assisted {MISO}
  systems,'' \emph{IEEE Trans. Wireless Commun.}, DOI:10.1109/TWC.2020.2986438,
  2020.

\bibitem{di2019smart}
M.~Di~Renzo, M.~Debbah, D.-T. Phan-Huy, A.~Zappone, M.-S. Alouini, C.~Yuen,
  V.~Sciancalepore, G.~C. Alexandropoulos, J.~Hoydis, H.~Gacanin \emph{et~al.},
  ``Smart radio environments empowered by reconfigurable {AI} meta-surfaces: An
  idea whose time has come,'' \emph{EURASIP J. Wireless Commun. Netw.}, vol.
  2019, no.~1, pp. 1--20, 2019.

\bibitem{jamali2019intelligent}
V.~Jamali, A.~M. Tulino, G.~Fischer, R.~M{\"u}ller, and R.~Schober,
  ``Intelligent reflecting and transmitting surface aided millimeter wave
  massive {MIMO},'' \emph{arXiv preprint arXiv:1902.07670}, 2019.

\bibitem{bjornson2019intelligent}
E.~Bj{\"o}rnson, {\"O}.~{\"O}zdogan, and E.~G. Larsson, ``Intelligent
  reflecting surface versus decode-and-forward: How large surfaces are needed
  to beat relaying?'' \emph{IEEE Wireless Commun. Lett.}, vol.~9, no.~2, pp.
  244--248, 2019.

\bibitem{jung2020performance}
M.~Jung, W.~Saad, Y.~Jang, G.~Kong, and S.~Choi, ``Performance analysis of
  large intelligent surfaces {(LISs)}: Asymptotic data rate and channel
  hardening effects,'' \emph{IEEE Trans. Wireless Commun.}, vol.~19, no.~3, pp.
  2052--2065, 2020.

\bibitem{zhang2020sr}
Q.~Zhang, Y.-C. Liang, and H.~V. Poor, ``Symbiotic radio: A new application of
  largeintelligent surface/antennas {(LISA)},'' \emph{Proc. of IEEE Wireless
  Commun. Netw. Conf. (WCNC)}, May 2020.

\bibitem{zhang2020large}
------, ``Large intelligent surface/antennas {(LISA)} assisted symbiotic radio
  for {IoT} communications,'' \emph{arXiv preprint arXiv:2002.00340}, 2020.

\bibitem{bharadia2015backfi}
D.~Bharadia, K.~R. Joshi, M.~Kotaru, and S.~Katti, ``{BackFi}: High throughput
  {WiFi} backscatter,'' in \emph{Proc. ACM SIGCOMM}, vol.~45, no.~4.\hskip 1em
  plus 0.5em minus 0.4em\relax London, United Kingdom: ACM, Aug. 2015, pp.
  283--296.

\bibitem{darsena2017modeling}
D.~Darsena, G.~Gelli, and F.~Verde, ``Modeling and performance analysis of
  wireless networks with ambient backscatter devices,'' \emph{IEEE Trans.
  Commun.}, vol.~65, no.~4, pp. 1797--1814, Apr. 2017.

\bibitem{smida2017reflectfx}
B.~Smida and S.~Khaledian, ``{ReflectFX}: In-band full-duplex wireless
  communication by means of reflected power,'' \emph{IEEE Trans. Commun.},
  vol.~65, no.~5, pp. 2207--2219, 2017.

\bibitem{yang2018optimal}
G.~Yang, D.~Yuan, Y.-C. Liang, R.~Zhang, and V.~C. Leung, ``Optimal resource
  allocation in full-duplex ambient backscatter communication networks for
  wireless-powered {IoT},'' \emph{IEEE Internet Things J.}, vol.~6, no.~2, pp.
  2612--2625, 2018.

\bibitem{xiao2019resource}
S.~Xiao, H.~Guo, and Y.-C. Liang, ``Resource allocation for full-duplex-enabled
  cognitive backscatter networks,'' \emph{IEEE Trans. Wireless Commun.},
  vol.~18, no.~6, pp. 3222--3235, 2019.

\bibitem{otto2006system}
C.~Otto, A.~Milenkovic, C.~Sanders, and E.~Jovanov, ``System architecture of a
  wireless body area sensor network for ubiquitous health monitoring,''
  \emph{J. Mobile Multimedia}, vol.~1, no.~4, pp. 307--326, 2006.

\bibitem{ricquebourg2006smart}
V.~Ricquebourg, D.~Menga, D.~Durand, B.~Marhic, L.~Delahoche, and C.~Loge,
  ``The smart home concept: our immediate future,'' in \emph{Proc. of IEEE Int.
  Conf. E-Learn. Ind. Electron. (ICELIE)}.\hskip 1em plus 0.5em minus
  0.4em\relax Hammamet, Tunisia: IEEE, Dec. 2006, pp. 23--28.

\bibitem{messer2006environmental}
H.~Messer, A.~Zinevich, and P.~Alpert, ``Environmental monitoring by wireless
  communication networks,'' \emph{Science}, vol. 312, no. 5774, pp. 713--713,
  2006.

\bibitem{he2019cascaded}
Z.-Q. He and X.~Yuan, ``Cascaded channel estimation for large intelligent
  metasurface assisted massive {MIMO},'' \emph{IEEE Wireless Commun. Lett.},
  vol.~9, no.~2, pp. 210--214, 2019.

\bibitem{chen2019channel}
J.~Chen, Y.-C. Liang, H.~V. Cheng, and W.~Yu, ``Channel estimation for
  reconfigurable intelligent surface aided multi-user {MIMO} systems,''
  \emph{arXiv preprint arXiv:1912.03619}, 2019.

\bibitem{taha2019enabling}
A.~Taha, M.~Alrabeiah, and A.~Alkhateeb, ``Enabling large intelligent surfaces
  with compressive sensing and deep learning,'' \emph{arXiv preprint
  arXiv:1904.10136}, 2019.

\bibitem{mirza2019channel}
J.~Mirza and B.~Ali, ``Channel estimation method and phase shift design for
  reconfigurable intelligent surface assisted mimo networks,'' \emph{arXiv
  preprint arXiv:1912.10671}, 2019.

\bibitem{goldsmith2005wireless}
A.~Goldsmith, \emph{Wireless communications}.\hskip 1em plus 0.5em minus
  0.4em\relax Cambridge university press, 2005.

\bibitem{Larsson-weave}
{L. V. d. Perre, E. G. Larsson, F. Tufvesson, L. D. Strycker, E. Björnson, and
  O. Edfor}, ``{RadioWeaves for efficient connectivity: analysis andimpact of
  constraints in actual deployments},'' \emph{{ arXiv preprint
  arXiv:2001.05779}}, 2020.

\end{thebibliography}

\end{document}